\documentclass[twocolumn,nofootinbib,amsmath,amssymb,a4paper]{revtex4}

\usepackage{graphicx}
\usepackage{dcolumn}
\usepackage{bm}

\begin{document}

\title{Theoretical issues with $\Delta S$ in three-body $B$ decays}

\author{Hai-Yang Cheng}
 \email{phcheng@phys.sinica.edu.tw}
\affiliation{%
Institute of Physics, Academia Sinica, Taipei, Taiwan, 115, R.O.C.\\
}%

\begin{abstract}
Theoretical uncertainties with the time-dependent $CP$ asymmetries
$\Delta S$ and $A_{\rm CP}$ in $B^0\to K^+K^-K_S$, $K_SK_SK_S$ and
$K_S\pi^0\pi^0$ decays are discussed. In order to have a reliable
estimate of $CP$ asymmetries, it is very crucial to understand the
underlying mechanism for the nonresonant background which
dominates the $KKK$ modes.
\end{abstract}

\maketitle

\section{Introduction}
Time-dependent mixing-induced and direct $CP$ asymmetries $S_f$
and $A_f$, respectively, in charmless three-body decays of the $B$
mesons have been measured by BaBar and Belle for $K_SK_SK_S$ and
$K^+K^-K^0$ final states. In addition, BaBar has also measured
time-dependent $CP$ asymmetries for $K^+K^-K_L$ and
$K_S\pi^0\pi^0$ and performed a Dalitz plot analysis for
$K^+K^-K^0$.

For $CP$ eigenstates, the measured mixing-induced $CP$ asymmetry
$S_f$ can be used to define an effective $\sin 2\beta$ via
 \begin{eqnarray}
 S_f\equiv -\eta_f\sin 2\beta_{\rm eff}
 \end{eqnarray}
with $\eta_f$ being the $CP$ eigenvalue of the final state $f$.
While $K_SK_SK_S$, $K_SK_SK_L$ and $K_S\pi^0\pi^0$ have fixed
$CP$-parities with $\eta_f=1,-1,1$ respectively, $K^+K^-K_S$ is an
admixture of $CP$-even and $CP$-odd components, rendering its CP
analysis more complicated. By excluding the major $CP$-odd
contribution from $\phi K_S$, the 3-body $K^+K^-K_S$ final state
is primarily $CP$-even. A measurement of the $CP$-even fraction
$f_+$ in the $B^0\to K^+K^-K_S$ decay yields
$f_+=0.89\pm0.08\pm0.06$ by BaBar \cite{BaBarSf2} and
$0.93\pm0.09\pm0.05$ by Belle \cite{BelleSf3}, while the CP-odd
fraction in $K^+K^-K_L$ is measured to be
$f_-=0.92\pm0.33^{+0.13}_{-0.14}\pm0.10$ by BaBar
\cite{BaBarKKKL}. Hence, while $\eta_f=1$ for the $K_SK_SK_S$
mode, $\eta_f=2f_+-1$ for $K^+K^-K_S$ and $\eta_f=-(2f_--1)$ for
$K^+K^-K_L$.

The world average of $\sin 2\beta$ measured in the decays
$B\to\phi_{c\bar c}K_S$  with $\phi_{c\bar c}$ being a charmonium
is $\sin 2 \beta_{\phi_{c\bar c} K_S}=0.675\pm0.026$ \cite{HFAG}.
However, the time-dependent {\it CP} asymmetries in the $b\to
sq\bar q$ penguin-induced two-body decays such as $B^0\to
(\phi,\omega,\pi^0,\eta',f_0)K_S$ measured by BaBar and Belle
indicate $\sin 2\beta_{\rm eff}=0.53\pm0.05$ \cite{HFAG}. This
seems to imply a 2.6 $\sigma$ deviation from the expectation of
the standard model (SM) where $CP$ asymmetry in all
above-mentioned modes should be equal to $\sin 2
\beta_{\phi_{c\bar c} K_S}$ \cite{HFAG} with a small deviation
{\it at most} ${\cal O}(0.1)$ \cite{LS}.

The measured results of $\sin 2\beta_{\rm eff}$ for
$K^+K^-K_{S,L}$, $K_SK_SK_S$ and $K_S\pi^0\pi^0$ are listed in
Table IV below. In order to see if the current measurements of the
deviation of $\sin 2\beta_{\rm eff}$ in these modes from $\sin 2
\beta_{\phi_{c\bar c} K_S}$ signal New Physics in $b\to s$
penguin-induced modes, it is of great importance to examine and
estimate how much of the deviation of $\sin 2\beta_{\rm eff}$ is
allowed in the SM. One of the major uncertainties in the dynamic
calculations lies in the hadronic matrix elements which are
nonperturbative in nature. One way to circumvent this difficulty
is to impose SU(3) flavor symmetry \cite{Grossman03,Engelhard} or
isospin and U-spin symmetries \cite{Gronau} to constrain the
relevant hadronic matrix elements. While this approach is model
independent in the symmetry limit, deviations from that limit can
only be computed in a model dependent fashion.  In addition, it
may have some weakness as discussed in \cite{Engelhard}.

As mentioned above, BaBar has performed a Dalitz plot analysis for
$B^0\to K^+K^-K^0$ decays and found that it is dominated by
$S$-wave nonresonant contributions with fraction $91\pm19\%$
\cite{BaBarKpKmK0}. A large nonresonant component is also found in
$B^+\to K^+K^+K^-$ decays \cite{BaBarKpKpKm,BelleKpKpKm}. This is
a surprise because it is known that in three-body decays of $D$
mesons, the nonresonant background is at most 10\%. Hence, it is
very important to take into account the large nonresonant effects
in any realistic model calculations.

\section{Formulism for charmless 3-body $B$ decays}

We consider the decay $\overline B^0\to K^+K^-\overline K^0$ as an
illustration. Under the factorization approach, its decay
amplitude consists of three distinct factorizable terms: (i) the
meson emission process, $\langle \overline B^0\to K^+\overline
K^0\rangle\times \langle 0\to K^-\rangle$, (ii) the transition
process, $\langle \overline B^0\to \overline K^0\rangle\times
\langle 0\to K^+K^-\rangle$, and (iii) the annihilation process
$\langle \overline B^0\to 0\rangle\times \langle 0\to
K^+K^-\overline K^0\rangle$, where $\langle A\to B\rangle$ denotes
a $A\to B$ transition matrix element.

For the kaon emission process, the amplitude induced by $V-A$
currents reads
 \begin{eqnarray}
 A_{\rm current-ind} &\equiv&\langle K^-(p_3)|(\bar s
 u)_{V-A}|0\rangle \nonumber \\ &\times& \langle \overline K^0 (p_1) K^+(p_2)|
 (\bar u b)_{V-A}|\overline B^0\rangle \nonumber\\
 &=& -\frac{f_K}{2}\large[2 m_3^2 r+(m_B^2-s_{12}-m_3^2) \omega_+
 \nonumber \\
 &+& (s_{23}-s_{13}-m_2^2+m_1^2) \omega_-\large],
 \end{eqnarray}
where $s_{ij}\equiv (p_i+p_j)^2$, $(\bar q_1q_2)_{V-A}\equiv \bar
q_1\gamma_\mu(1-\gamma_5)q_2$, and $r,~\omega_+$ and $\omega_-$
are three unknown form factors. In principle, one can apply heavy
meson chiral perturbation theory (HMChPT) to evaluate the form
factors $r,~\omega_+$ and $\omega_-$ \cite{LLW}. The relevant
diagrams are depicted in Fig. 1. However, this will lead to too
large nonresonant decay rates in disagreement with experiment
\cite{Cheng:2002qu}. A direct calculation indicates that the
branching ratio of $B^0\to K^+K^-K^0$ arising from the
current-induced process alone is already at the level of $77\times
10^{-6}$ which substantially exceeds the measured total branching
ratio of $25\times 10^{-6}$ \cite{HFAG}. The issue has to do with
the applicability of HMChPT. In order to apply this approach, two
of the final-state pseudoscalars have to be soft. The momentum of
the soft pseudoscalar should be smaller than the chiral symmetry
breaking scale of order 1 GeV. For 3-body charmless $B$ decays,
the available phase space where chiral perturbation theory is
applicable is only a small fraction of the whole Dalitz plot.
Therefore, it is not justified to apply chiral and heavy quark
symmetries to a certain kinematic region and then generalize it to
the region beyond its validity.

To circumvent the aforementioned difficulty, we assume the
momentum dependence of the nonresonant amplitude in the
exponential form
 \begin{eqnarray} \label{eq:ADalitz}
  A_{\rm current-ind}=A_{\rm current-ind}^{\rm
  HMChPT}\,e^{-\alpha_{_{\rm NR}}
p_B\cdot(p_1+p_2)}e^{i\phi_{12}},
 \end{eqnarray}
so that the HMChPT results are recovered in the soft meson limit
$p_1,p_2\to 0$. That is, the nonresonant amplitude in the soft
meson region is described by HMChPT, but its energy dependence
beyond the chiral limit is governed by the exponential term
$e^{-\alpha_{_{\rm NR}} p_B\cdot(p_1+p_2)}$. The parameter
$\alpha_{_{\rm NR}}$ can be constrained by the decay
$B^-\to\pi^+\pi^-\pi^-$ \cite{CCS3body}.

\begin{figure}
 \includegraphics[width=0.35\textwidth]{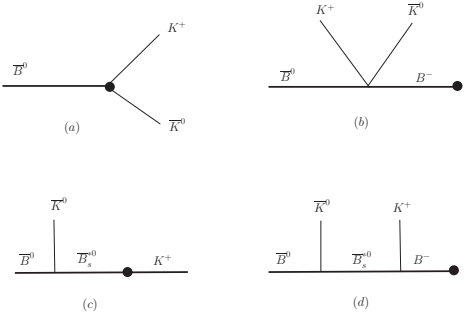}
    \caption[]{\small Point-like and pole diagrams responsible for the $\overline B^0\to K^+\overline K^0$
    matrix element of the current $\bar u\gamma_\mu(1-\gamma_5)b$,
    where the symbol $\bullet$ denotes an insertion of the current.}
\end{figure}

For the kaon emission process, there exist two more amplitudes,
one induced by currents and the other induced by scalar densities:
 \begin{eqnarray}
 A_1&=&\langle \overline K^0(p_3)|(\bar s
 d)_{V-A}|0\rangle \langle K^+ (p_2) K^-(p_3)|(\bar d b)_{V-A}|\overline B^0\rangle, \nonumber\\
 A_2&=&\langle \overline K^0(p_3)|\bar s
 d|0\rangle \langle K^+ (p_2) K^-(p_3)|\bar d b|\overline
 B^0\rangle.
 \end{eqnarray}
Although the 3-body matrix elements in $A_1$ and $A_2$ are OZI
suppressed, they do receive intermediate vector meson and scalar
pole contributions, respectively. For example,
 \begin{eqnarray}
 && \langle K^+(p_2)K^-(p_3)|(\bar db)_{V-A}|\overline B^0\rangle^R =
 \nonumber  \\  && \sum_{i}\frac{g^{{f_0}_i\to K^+K^-}}{m_{{f_0}_i}^2-s_{23}-i
 m_{{f_0}_i} }\langle {f_0}_i|(\bar
db)_{V-A}|\overline B^0\rangle,
 \end{eqnarray}
where ${f_0}_i$ denote the generic $f_0$-type scalar mesons,
${f_0}_i=f_0(980),f_0(1370),f_0(1500),X_0(1550),\cdots$.

For the transition amplitudes, we need to consider the two-body
matrix elements $\langle K^+K^-|V_\mu|0\rangle$ and $\langle
K^+K^-|\bar ss|0\rangle$. Both receive resonant and nonresonant
contributions. The first matrix element can be related to the kaon
electromagnetic (e.m.) form factors $F^{K^+K^-}_{em}$ and
$F^{K^0\bar K^0}_{em}$ for the charged and neutral kaons,
respectively. The nonresonant form factor can be constrained by
the asymptotic behavior implied by pQCD \cite{Brodsky}. The second
matrix element has the expression
 \begin{eqnarray} \label{eq:KKssme}
 \langle K^+(p_2) K^-(p_3)|\bar s s|0\rangle
&=& \sum_{i}\frac{m_{{f_0}_i} \bar f^s_{{f_0}_i} g^{{f_0}_i\to
K^+K^-}}{m_{{f_0}_i}^2-s_{23}-i
 m_{{f_0}_i}\Gamma_{{f_0}_i}} \nonumber \\  &+& f_s^{NR},
 \end{eqnarray}
where the scalar decay constant $\bar f_{{f_0}_i}^s$ is defined by
$\langle {f_0}_i|\bar s s|0\rangle=m_{{f_0}_i} \bar
f^s_{{f_0}_i}$.

It turns out that the nonresonant background in $B\to P_1P_2$
transition described by Eq. (\ref{eq:ADalitz}) it too small to
account for the experimental observation that $\overline B^0\to
K^+K^-\overline K^0$ is dominated by the nonresonant
contributions. This implies that the two-body matrix element e.g.
$\langle K\overline K|\bar ss|0\rangle$ induced by the scalar
density should have a large nonresonant component. In the absence
of first-principles calculation, we will use the decay mode
$\overline B^0\to K_SK_SK_S$ or $B^-\to K^-K_SK_S$ to fix
$f_s^{NR}$ \cite{CCS3body}.

The calculated branching ratios of resonant and nonresonant
contributions to $\overline B^0\to K^+K^-\overline K^0$ and
$K_S\pi^0\pi^0$ are summarized in Tables \ref{tab:KpKmK0} and
\ref{tab:K0pi0pi0}, respectively. The theoretical errors shown
there are from the uncertainties in (i) the parameter
$\alpha_{_{\rm NR}}$ which governs the momentum dependence of the
nonresonant amplitude [cf. Eq. (\ref{eq:ADalitz})], (ii) the
strange quark mass $m_s$, the form factor $F^{BK}_0$ and the
nonresonant contribution $f_s^{NR}$ constrained by the $K_SK_SK_S$
rate, and (iii) the unitarity angle $\gamma$. We see that the
predicted rates for resonant and nonresonant components are
consistent with experiment. The nonresonant background arises
dominantly from the transition process (90\%) via the scalar
density induced vacuum to $K\bar K$ transition, namely, $\langle
K^+K^-|\bar ss|0\rangle$, and slightly from the current-induced
process (3\%). In the $K_S\pi^0\pi^0$ mode, the nonresonant
fraction is found to be of order 40\%. Recall that large
nonresonant contributions to the decays $B^-\to K^-\pi^+\pi^-$ and
$\overline  B^0\to\overline{K}^0\pi^+\pi^-$, at the level of
$(35-40)\%$, have been reported by Belle \cite{BelleKpipi}.

\begin{table}[h]
\caption{Branching ratios (in units of $10^{-6}$) of resonant and
nonresonant (NR) contributions to $\overline B^0\to
K^+K^-\overline K^0$. Theoretical errors correspond to the
uncertainties in (i) $\alpha_{\rm NR}$, (ii) $m_s$, $F^{BK}_0$ and
$\sigma_{\rm NR}$, and (iii) $\gamma$. Experimental results are
taken from \cite{BaBarKpKmK0}.} \label{tab:KpKmK0}
\begin{center}
\begin{tabular}{l l  l} \hline
 Decay mode~~ & BaBar \cite{BaBarKpKmK0}  & Theory  \\ \hline
 $\phi \overline K^0$ & $3.07\pm0.45$ & $2.6^{+0.0+0.5+0.0}_{-0.0-0.4-0.0}$ \\
 $f_0(980)\overline K^0$ & $5.31\pm2.19$ & $5.8^{+0.0+0.1+0.0}_{-0.0-0.5-0.0}$ \\
 $X_0(1550)K^-$ & $0.98\pm0.44$ & $0.93^{+0.00+0.16+0.00}_{-0.00-0.15-0.00}$ \\
 NR & $21.7\pm5.1$ & $14.4^{+0.4+2.1+0.1}_{-0.5-1.5-0.1}$ \\
\hline
 total & $23.8\pm2.0\pm1.6$  &  $17.1^{+0.3+3.6+0.1}_{-0.3-2.9-0.1}$ \\
 \hline
\end{tabular}
\end{center}
\end{table}

\begin{figure*}
 \includegraphics[width=0.3\textwidth]{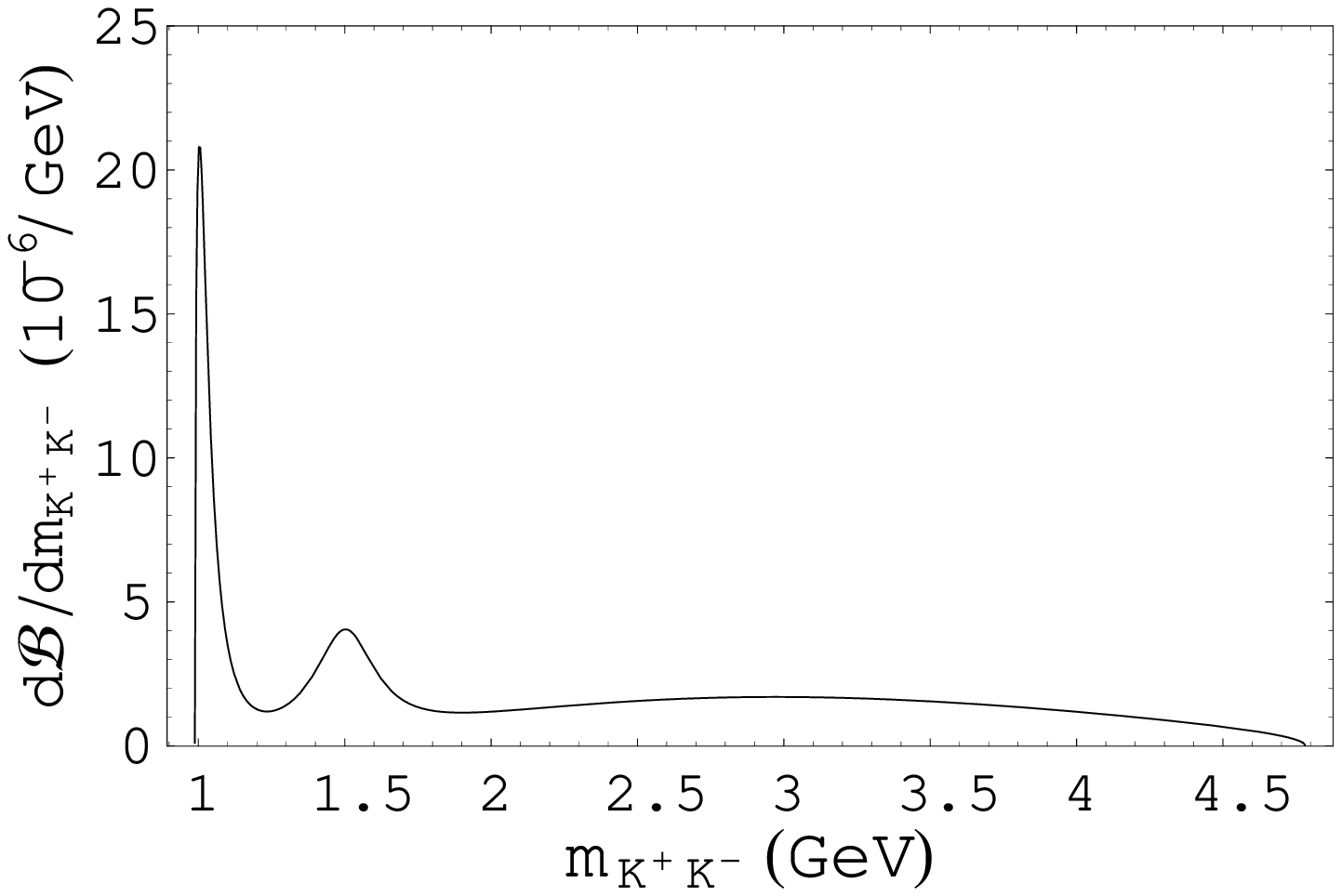}
  \hspace{0.2cm}\includegraphics[width=0.305\textwidth]{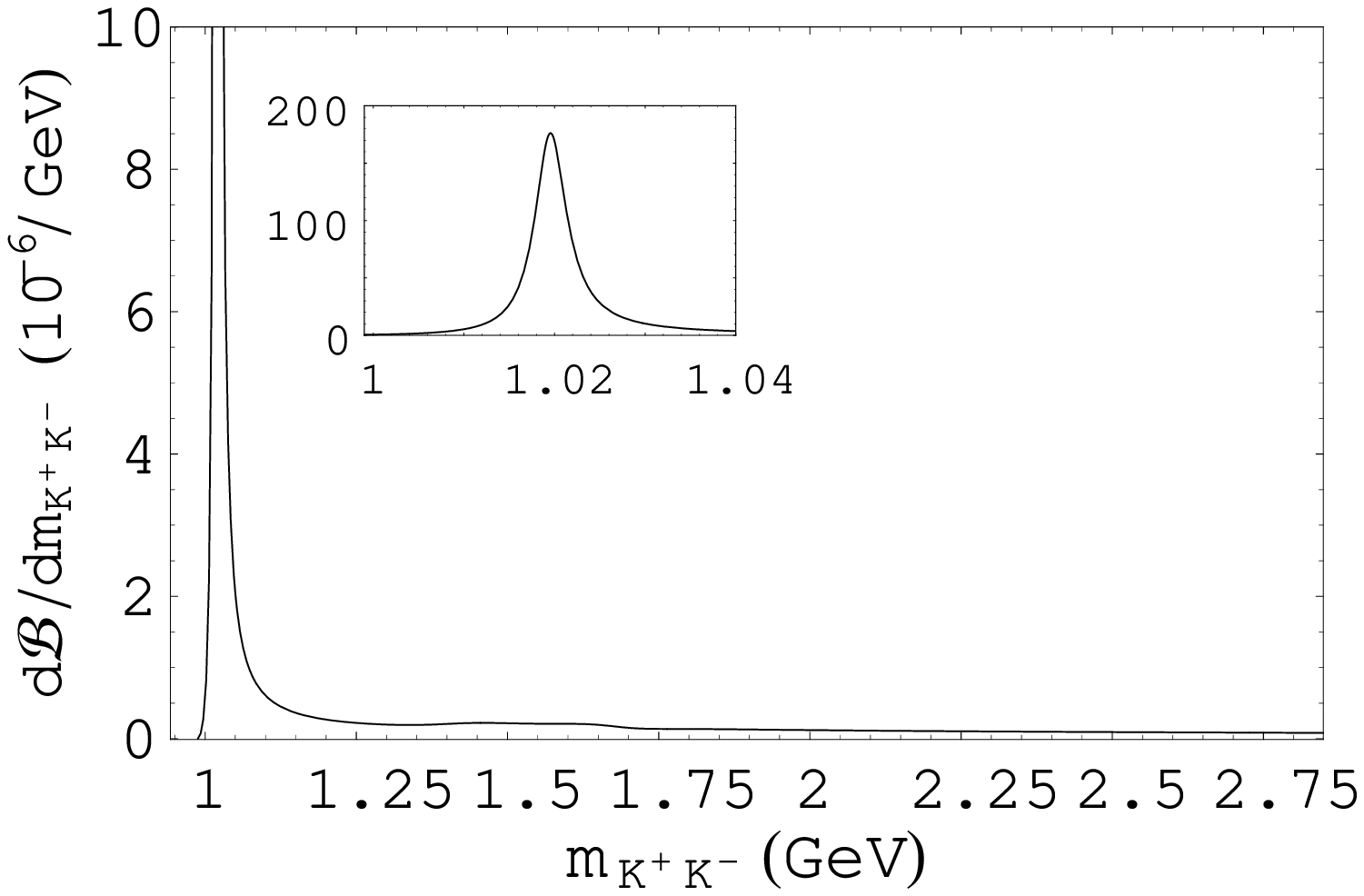}
  \hspace{0.2cm}\includegraphics[width=0.29\textwidth]{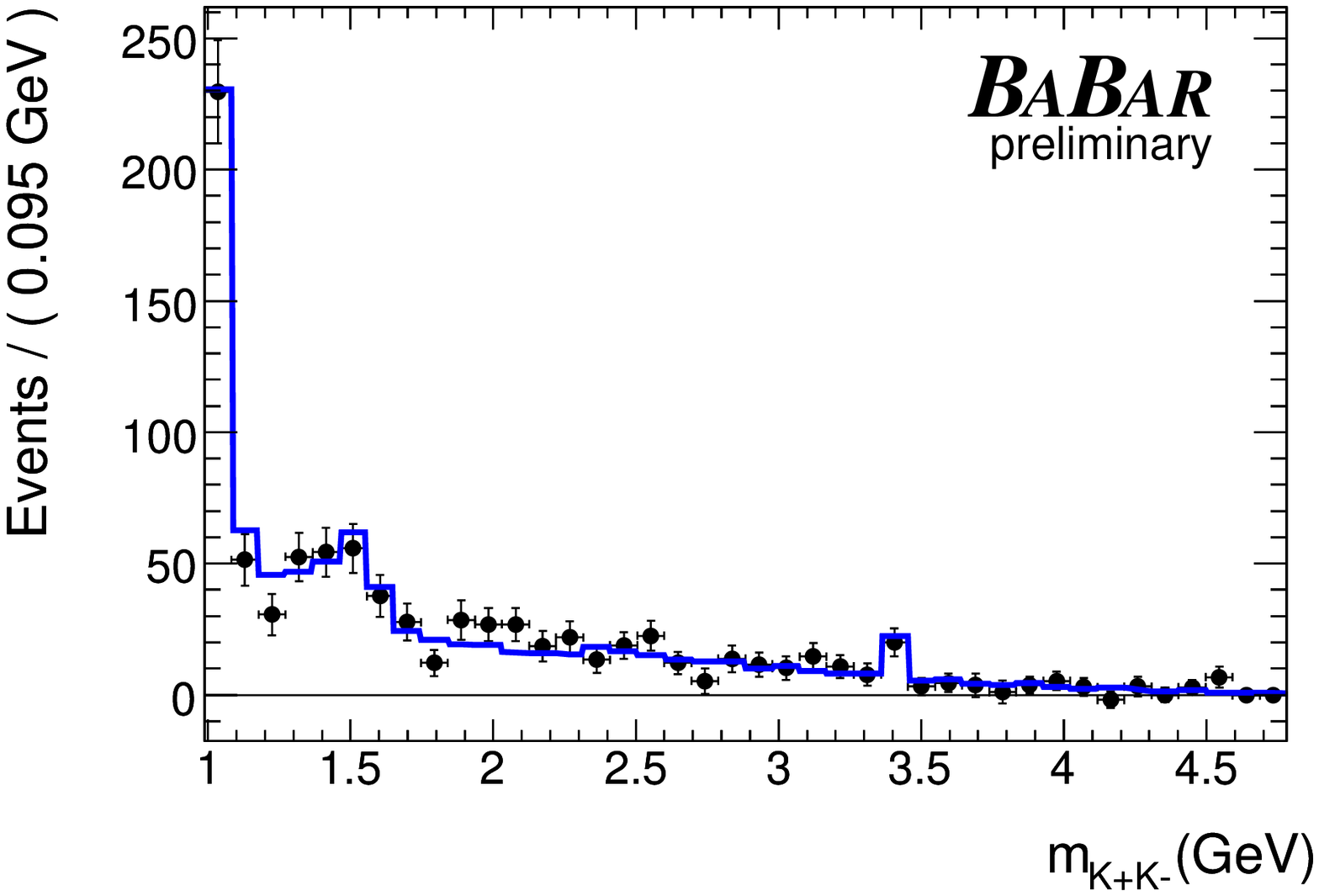}
    \centerline{(a)
              \hspace{5.3cm}
              (b)
              \hspace{5.5cm}
              (c)
              }
    \caption{\small The $K^+ K^-$ mass spectra for
    $\overline B {}^0\to K^+ K^- K_S$ decay from (a) $CP$-even and (b) $CP$-odd contributions.
    The insert in (b) is for the $\phi$ region. The
full $K^+K^-K_S$ spectrum, which is the sum of $CP$-even and
$CP$-odd parts, measured by BaBar \cite{BaBarKpKmK0} is depicted
in (c). }\label{fig:rates}
\end{figure*}

\begin{table}[h]
\caption{Same as Table \ref{tab:KpKmK0} except for the $\overline
K^0\pi^0\pi^0$ mode.} \label{tab:K0pi0pi0}
\begin{center}
\begin{tabular}{l   l} \hline
 Decay mode~~ &  Theory  \\ \hline
 $f_0(980)\overline K^0$ & $3.8^{+0.0+2.0+0.0}_{-0.0-0.4-0.0}$ \\
 $\overline K^{*0}\pi^0$ & $0.55^{+0.00+0.16+0.00}_{-0.00-0.13-0.00}$ \\
 $\overline K^{*0}_0(1430)\pi^0$ &
 $2.3^{+0.0+0.8+0.0}_{-0.0-0.6-0.0}$\\
 NR & $5.1^{+0.1+1.8+0.0}_{-0.1-1.0-0.0}$  \\
\hline
 total & $12.5^{+0.0+3.9+0.1}_{-0.0-2.9-0.1}$ \\
 \hline
\end{tabular}
\end{center}
\end{table}

The $K^+K^-K_S$ mode is an admixture of $CP$-even and $CP$-odd
components. By excluding the major $CP$-odd contribution from
$\phi K_S$, the 3-body $K^+K^-K_S$ final state is primarily
$CP$-even. The $K^+K^-$ mass spectra of the $\overline B {}^0\to
K^+ K^- K_S$ decay from $CP$-even and $CP$-odd contributions are
shown in Fig.~\ref{fig:rates}.  For the $CP$-even spectrum, there
are peaks at the threshold and $m_{K^+K^-}=1.5$ GeV
region.\footnote{In our previous work \cite{CCSKKK} we have argued
that the spectrum should have a peak at the large $m_{K^+ K^-}$
end. This is because we have introduced an additional nonresonant
contribution to the $\omega_-$ parameter parametrized as
$\omega^{NR}_-=\kappa\, \frac{2p_B\cdot p_2}{s^2_{12}}$ and
employed the $B^-\to D^0 K^0 K^-$ data and applied isospin
symmetry to the $\overline B\to K\overline K$ matrix elements to
determine the unknown parameter $\kappa$. Since this nonresonant
term favors a small $m_{K^+ K_S}$ region, a peak of the spectrum
at large $m_{K^+K^-}$ is thus expected. However, such a bump is
not seen experimentally \cite{BaBarKpKmK0} [see also Fig.
\ref{fig:rates}(c)]. In this work we will no longer consider this
term.}
The threshold enhancement arises from the $f_0(980) K_S$ and the
nonresonant $f_S^{K^+K^-}$ contributions. For the $CP$-odd
spectrum, the peak on the lower end corresponds to the $\phi K_S$
contribution, which is also shown in the insert. The small middle
hump in Fig.~\ref{fig:rates}(a) comes from the interference
between $b\to s$ and $b\to u$ amplitudes. The $b\to u$ transition
is governed by the current-induced process $\langle \overline
B^0\to K^+\overline K^0\rangle\times \langle 0\to K^-\rangle$.
From Eq. (\ref{eq:ADalitz}) it is clear that the $b\to u$
amplitude prefers a small invariant mass of $K^+$ and $\overline
K^0$ and hence a large invariant mass of $K^+$ and $K^-$. In
contrast, the $b\to c$ amplitude prefers a small $s_{23}$. The
interference results to a minor hump shown in
Fig.~\ref{fig:rates}(a).

\section{Decay rates and $CP$ asymmetries}

\begin{table}[t]
\caption{Branching ratios for $\overline B {}^0\to K^+ K^-
K_S,\,K_S K_S K_S,\,K_S K_S K_L\,K_S\pi^0\pi^0$ decays and the
fraction of $CP$-even contribution to $\overline B^0\to
K^+K^-K_S$. The branching ratio of $CP$-odd $K^+K^-K_S$ with $\phi
K_S$ excluded is shown in parentheses. Results for
$(K^+K^-K_L)_{CP\pm}$ are identical to those for
$(K^+K^-K_S)_{CP\mp}$.  Experimental results are taken from
\cite{HFAG}.} \label{tab:BrKKK}
\begin{ruledtabular}
\begin{tabular}{c r r}
Final State &${\cal B}(10^{-6})_{\rm theory}$ &${\cal B}(10^{-6})_{\rm expt}$ \\
\hline
 $K^+ K^- K_S$
       & $8.54^{+0.13+1.82+0.06}_{-0.14-1.45-0.06}$
       & $12.4\pm1.2$ \\
 $(K^+ K^- K_S)_{CP+}$
       & $6.97^{+0.04+1.37+0.04}_{-0.04-1.12-0.04}$
       &  \\
 $(K^+ K^- K_S)_{CP-}$
       & $1.57^{+0.09+0.46+0.02}_{-0.10-0.32-0.02}$
       &  \\
       & $(0.14^{+0.06+0.14+0.01}_{-0.06-0.06-0.01})$
       &  \\
 $K_S K_S K_S$
       & input
       & $6.2\pm0.9$ \\
 $K_S K_S K_L$
       & $6.06^{+0.11+0.61+0.02}_{-0.08-0.69-0.02}$
       & $<14$
       \\
 $K_S\pi^0\pi^0$
       & $6.24^{+0.01+1.96+0.03}_{-0.02-1.45-0.04}$
       &
       \\
 \end{tabular}
\end{ruledtabular}
\end{table}

\begin{table*}
\caption{Mixing-induced and direct $CP$ asymmetries $\sin
2\beta_{\rm eff}$ (top) and $A_f$ (in $\%$, bottom), respectively,
in $B^0\to K^+K^-K_S$, $K_SK_SK_S$ and $K_S\pi^0\pi^0$ decays.
Results for $(K^+K^-K_L)_{CP\pm}$ are identical to those for
$(K^+K^-K_S)_{CP\mp}$. Experimental results are taken from
\cite{HFAG}.} \label{tab:ASKKK}
\begin{ruledtabular}
\begin{tabular}{l r r}
 Final state & $\sin 2\beta_{\rm eff}$  & Expt.  \\
 \hline
 $(K^+K^-K_S)_{\phi K_S~{\rm excluded}}$
            & $0.721^{+0.000+0.001+0.009}_{-0.001-0.001-0.020}$
            & $0.58\pm0.13^{+0.12}_{-0.09}$
            \\
 $(K^+K^-K_S)_{CP+}$
            & $0.726^{+0.002+0.007+0.008}_{-0.002-0.004-0.019}$
            &
            \\
 $(K^+K^-K_L)_{\phi K_L~{\rm excluded}}$
             & $0.721^{+0.000+0.001+0.009}_{-0.001-0.001-0.020}$
            & $0.58\pm0.13^{+0.12}_{-0.09}$
            \\
 $K_SK_SK_S$
            & $0.719^{+0.000+0.000+0.008}_{-0.000-0.000-0.019}$
            & $0.51\pm0.21$
            \\
 $K_SK_SK_L$
            & $0.718^{+0.000+0.000+0.008}_{-0.000-0.000-0.019}$
            &
            \\
 $K_S\pi^0\pi^0$
            & $0.729^{+0.001+0.001+0.009}_{-0.001-0.001-0.019}$
            & $-0.84\pm0.71\pm0.08$
            \\
 \hline
  &$A_f(\%)$  &Expt. \\
 \hline
 $(K^+K^-K_S)_{\phi K_S~{\rm excluded}}$
            & $-4.11^{+1.22+0.41+0.35}_{-0.91-0.42-0.30}$
            & $-15\pm9$
            \\
 $(K^+K^-K_S)_{CP+}$
            & $-4.37^{+1.30+0.45+0.37}_{-1.00-0.46-0.31}$
            &
            \\
 $(K^+K^-K_L)_{\phi K_L~{\rm excluded}}$
            & $-4.11^{+1.22+0.41+0.35}_{-0.91-0.42-0.30}$
            & $-15\pm9$
            \\
 $K_SK_SK_S$
            & $0.69^{+0.01+0.01+0.05}_{-0.01-0.01-0.06}$
            & $23\pm15$
            \\
 $K_SK_SK_L$
            & $0.90^{+0.00+0.00+0.06}_{-0.00-0.02-0.08}$
            &
            \\
 $K_S\pi^0\pi^0$
            & $0.32^{+0.05+0.02+0.02}_{-0.03-0.01-0.03}$
            & $-27\pm52\pm13$
            \\
 \end{tabular}
\end{ruledtabular}
\end{table*}

Results for the decay rates and $CP$ asymmetries in $\overline B
{}^0\to K^+ K^- K_{S(L)},\,K_S K_S K_{S(L)}$ are exhibited in
Table~\ref{tab:BrKKK} and Table~\ref{tab:ASKKK}, respectively. We
see that the predicted branching ratios are consistent with the
data within the theoretical and experimental errors. From Table
\ref{tab:ASKKK} we can calculate the deviation of the
mixing-induced $CP$ asymmetry in $B^0\to K^+K^-K_S$ and
$K_SK_SK_S$ from that measured in $B\to \phi_{c\bar c}K_S$,
namely, $\Delta \sin 2\beta_{\rm eff}\equiv \sin 2\beta_{\rm
eff}-\sin 2 \beta_{\phi_{c\bar c}K_S}$.  Using  the fitted CKM's
$\sin2\beta=0.695^{+0.018}_{-0.016}$ \cite{CKMfitter} we obtain
 \begin{eqnarray}
 \label{eq:DeltaS}
 \Delta \sin 2\beta_{K^+K^-K_S}&=&0.026^{+0.022}_{-0.030}, \nonumber \\
 \Delta \sin 2\beta_{K_SK_SK_S}&=&0.024^{+0.020}_{-0.025},
 \nonumber\\
 \Delta \sin 2\beta_{K_S\pi^0\pi^0}&=&0.034^{+0.020}_{-0.025}.
 \end{eqnarray}
It should be stressed that despite the presence of color-allowed
tree contributions in $B^0\to K^+K^-K_{S(L)}$,  the deviation of
the mixing-induced $CP$ asymmetry in this penguin-dominated mode
from that measured in $B\to \phi_{c\bar c}K_S$ is very similar to
that of the $K_SK_SK_S$ mode. However, direct $CP$ asymmetry of
the former, being of order 4\%, is more prominent than the latter.

\newpage

\begin{acknowledgments}
 I'm grateful to Chun-Khiang Chua and  Amarjit Soni for fruitful
 collaboration.
\end{acknowledgments}

\end{document}